\documentstyle{article}

\begin{document}

\title{Nolinear waves, differential resultant, computer algebra and \\
completely integrable dynamical systems}

\author{N.~A.~Kostov, Z.~T.~Kostova}
\maketitle

\begin{center}
Institute of Electronics,Bulgaian Academy of Sciences, \\
Blvd. Tsarigradsko shosse 72, Sofia 1784, Bulgaria  \\
E-mail: nakostov@ie.bas.bg
\end{center}

\begin{abstract}
The hierarchy of integrable equations are considered. The dynamical
approach to theory of nonlinear waves is proposed. The special solutions
(nonlinear waves) of considered equations are derived. We use powerful
methods of computer algebra such as differential resultant and others.
\end{abstract}

\section{Introduction} \label{intro}

The paper is organized as follows: section 1 describes the
basic facts of hierarchy of nonlinear equations, Baker-Akhiezer (BA) function
theory for finite--gap integrationn method.
Particular solutions in terms of generalized Novikov, Lam\'e, Hermite type
polynomials are presented.

\section{ Preliminaries }

We review in this section some basic facts about Baker--Akhieser
function which will be used in the sequel. For systematic treatments
and proofs, we refer the reader to \cite{Kr2}, \cite{Cher}.

\begin{eqnarray}
[\partial_{x}-U,\partial_{t}-V^{(N)}]=0, \\
U=U(x,t,\lambda),\qquad V^{(N)}=V^{(N)}(x,t,\lambda),
\label{ZqE}
\end{eqnarray}
where $V^{(N)}=V_{0}\lambda^N+V_{1}\lambda^{(N-1)}+\ldots V_N$.
The nonlinear equations in the form (\ref{ZqE}) are obtain as
compatibility condition of the following linear differential
equations
\begin{eqnarray}
\partial_{x}\Psi=U(x,t,\lambda)\Psi, \qquad
\partial_{t}\Psi=V_{t}^{(N)}(x,t,\lambda)\Psi.
\label{LS}
\end{eqnarray}
where $\Psi$ is a vector-collumn $(\Psi_{1} \ldots \Psi_{n})^{T}$.
The system (\ref{LS}) play an impotran role in the theory of
finite-gap integration method. The function $\Psi$, which is
general solution of the system (\ref{LS}), is meromorphic
function on the Riemann surface and have esential singularities
of prescribed form near fixed points on Riemann surface. This surface
is defined by
\begin{equation}
\mbox{det}(L(x,t,\lambda)-\mu\mbox{I})=0.
\end{equation}
We call finite-gap solutions of (\ref{ZqE}) these for which there
exists the meromorphic in $\lambda$ matrix function $L(x,t,\lambda)$
such that
\begin{eqnarray}
L_{t} = [V^{i}, L], \quad i=1,\ldots ,N \\
L_{x} = [U,L],\quad L=L_{0}\,\lambda^{N} +
L_{1}\,\lambda^{N-1} + \ldots L_{N} .
\end{eqnarray}

\section{Resultant of two differential operators}

Let us consider the following two differential operators
\begin{eqnarray}\label{A.1}
 A =\sum_{k=0}^n a_{k}D^{k}, \quad a_{n}\neq 0, \qquad
 B =\sum_{l=0}^m b_{l}D^{l}, \quad b_{m}\neq  0, \quad a_{k},
b_{l}\in  C^{k(l)}_{I}
\end{eqnarray}
where $D = d/dx$, $C^{k(l)}_{I}$ is the set of $k(l)$-th
differentiable functions on the interval $I$ of real variable $x$.
Right resultant of operators $A$ and $B$ $\mbox{RRes}\,(A,B)$  is called
the determinant of resultant matrix $R$ of degree $(m+n)$.
Analogically left resultant of operators $A$ and $B$ $\mbox{LRes}\,(A,B)$
is called the resultant matrix  $R^{*}$ of degree $(m+n)$, i.e.
$\mbox{RRes}\,(A,B) = \det (R)$,  $\mbox{LRes}\,(A,B) = \det (R^{*})$.
By definition $\mbox{LRes}\,(A,B)$   = $\mbox{RRes}\,(A^{*},B^{*})$,
where
$A^{*}$ and $B^{*}$ are the conjugated operators. The differential
resultant answers to the question when the operators $A$ and $B$ have
right(left) divisor i.e. when the overdetermined system $Ay=0$, $By=0$
(or $A^{*}y=0$, $B^{*}y=0$) has soluion. By construction to define right
resultant $\mbox{RRes}\,(A,B)$ we act with operators $D^{m-1},\ldots ,D$,
$D^{0}=1$, by left to $A$ and with operators $D^{n-1},\ldots, D$,
$D^{0}=1$. to $B$.  In this case the system $Ay=0$, $By=0$ has the
following form \begin{eqnarray}\label{A.2} \sum_{k=0}^{n+s}a_{k,s}y^{(k)}
= 0 , \qquad s = 0 \div  (m-1), \qquad \sum^{m+p}_{l=0} b_{l,p}y^{(l)} =
0, \qquad p = 0 \div  (n-1), \end{eqnarray} where the coefficients
$a_{k,s}$ and $b_{l,p}$ are computed by:  \begin{eqnarray}\label{A.3}
a_{k,s}= \sum_{i=0}^{s} \left({s\atop i}\right) a^{(s-i)}_{k-i},\qquad
b_{l,p} = \sum_{j=0}^{p}\left({p\atop j}\right) b^{(p-j)}_{l-j}.
\end{eqnarray}
The system (\ref{A.2}) is constructed by homogeneous linear algebraic
equations in terms of the following variables
$y^{(n+m-1)},\ldots, y'$, $y$. The matrix of coefficients of this system is the
right resultant of the matrix
$R$,  $\det (R)= \mbox{RRes}\,(A,B)$.
\begin{eqnarray}\label{A.4}
\mbox{RRes}\,(A,B) =  \left| \begin{array}{cccccc}
a_{n+m-1,m-1} & a_{n+m-2,m-1} & \ldots &\ldots &\ldots & a_{0,m-1}\\
0 & a_{n+m-1,m-2} & \ldots &\ldots &\ldots  & a_{0,m-1}\\
\ldots &\ldots &\ldots  &\ldots &\ldots &\ldots \\
0 &\ldots &\ldots  &a_{n,0} &\ldots &a_{0,0} \\
b_{n+m-1,n-1} & b_{n+m-2,n-1} & \ldots &\ldots &\ldots & b_{0,n-1}\\
0 & b_{n+m-2,n-2} & \ldots &\ldots &\ldots & b_{0,n-1}\\
\ldots &\ldots &\ldots  &\ldots &\ldots &\ldots \\
0 &\ldots &\ldots  &b_{m,0} &\ldots &b_{0,0} \end{array} \right| .
\end{eqnarray}
If the system $Ay=0$, $ By=0$ is consistent then
$\det(R)=\mbox{RRes}\, (A,B)=0$, in the opposite case the system is
not consistent.

For applications of differential resultant see for example \cite{BerZ}.
The most important of them are the following: the cirterion of existing the
greater right and left divisor of the system of operators; the criteion
of concictency of the system of linear ordinary differential equations with
one unknown function; the criterion of commutation of two linear differential
operators; he criterion of existing polynomial and exponential solutions of
linear differential equtions with variable coefficients; the criterion of
factrorization of operators of degree $n$ into products of operators
of degrees $n-1$ and $1$.

\medskip
{\bf Example.} Let us introduce the following equations,
\begin{eqnarray}
& L1 \equiv Ay = x^{2}y'' + xy' - (x^{2} + 1/4 )y = 0,  & \\
& L2 \equiv By = 2xy'' + (3-4x)y' + (2x-3)y = 0 . &
\end{eqnarray}
Using  (\ref{A.4}) we obtain
$$
\mbox{RRes}\,(A,B)   = \left| \begin{array}{cccc}
x^{2} & 3x & -x^{2}+3/4 & -2x \\
0 & x^2 & x & - x^2 - 1/4 \\
2x & 5x-4 & 2x-7 & 2 \\
0 & 2x & 3 - 4x & 2x -3 \end{array} \right|
$$

\medskip
{\bf  Problem 1.} Let us define two LODE:
\begin{eqnarray}
Ay =\sum_{k=0}^n a_{k}D^{k}y =0 , \qquad a_{n}\neq  0, \qquad
 By =\sum_{l=0}^m b_{l}D^{l}y =0 , \qquad b_{m}\neq  0.
 a_{k},
b_{l}\in  C^{k(l)}_{I}
\end{eqnarray}
Find differential resultant of operators $A$ and $B$.

\medskip
{\bf Algorithm 1.} We solve problem 1 by the procedure
DIFRESULT$(a,n,b,m,x)$;
\begin{description}
\item[Input]
\item[$a$] is the array of coefficients of differential operator $A$;
\item[$n$] is the degree of differential operator $A$;
\item[$b$] is the array of coefficients of differential operator $B$;
\item[$m$] is the degree of differential operator $B$;
\item[$x$] is independent variable.
\item[Output:]{}
\end{description}

Using (\ref{A.3}) with given coefficients of differential operator  $A$
and $B$ we compute the ellements of the resultant matrix , after that we
we find the determinant. The output is $\mbox{RRes}\,(A,B) =
\det (R)$, where $R$ is the resultant matrix.

\section{The generalized Riccati equation}

{\bf 4.2.1.\, Prime right divisor of operator $L$.}

Let us define the differential operator $L$ of degree $n$:
$$
L =\sum_{k=0}^n a_{k}D^{k}, \qquad a_{n}\neq  0, a_{k}\in  C^{k}_{I} .
$$
A necessary and sufficient condition for the following factorization:
$$
L = L_{2}(D - \alpha ),\qquad \mbox{ord}\,L_{2} = n - 1
$$
is given by
$\mbox{RRes}\,(L,D-\alpha ) = 0$.
\begin{eqnarray}\label{AA.4}
\mbox{RRes}\,(L,D-\alpha) =  \left| \begin{array}{cccccc}
a_{n} & a_{n-1} & \ldots &a_{2} &a_{1} & a_{0}\\
1 & -\alpha & \ldots & -\left( {n-1 \atop n-3 } \right) \alpha^{(n-3)} &
-\left( {n-1 \atop n-2 } \right) \alpha^{(n-2)} & -\alpha^{(n-1)} \\
\ldots &\ldots & \ldots & \ldots & \ldots & \ldots  \\
0 & 0 & \ldots & 1 & -\alpha & -\alpha' \\
0 & 0 & \ldots & 0 &  1 & -\alpha
\end{array} \right| = 0.
\end{eqnarray}
Then we find the following generalized Riccati equation of first kind
$N(\alpha )$ of degree $(n-1)$:
$$
N(\alpha ) = L(\alpha ) + M(\alpha ),$$
where
$$
L(\alpha ) =\sum_{k=0}^n a_{k}\alpha ^{k},
$$
and $L(\alpha )$ is called he pseudo-characteristic Riccati equation of
first kind
$$
M(\alpha ) =\sum_{k=0}^n a_{k}M_{k-1}, \qquad M_{k-1} = (D + \alpha
)^{(k-1)} - \alpha ^{(k-1)}
$$
and $M(\alpha )$ is called the reduced Riccati equation of first kind.
When $n=2$ the generalized Riccati equation of first kind have the form
$$
N(\alpha ) = a_{2}\alpha ' + a_{2}\alpha ^{2} + a_{1}\alpha  + a_{0}
$$

{\bf 4.2.2. Prime left divisor of operator $L$.}
Let us define the following differential operator $L$ of degree $n$:
$$
L =\sum_{k=0}^n a_{k}D^{k}, \qquad a_{n}\neq 0 , \quad a_{k}\in C^{k}_{I}
$$
To find factorization of following type,
$$
L = L_{2}(D - \alpha ),\qquad \mbox{ ord} L_{2} = n - 1
$$
the necessary and sufficient condition is $\mbox{RRes}\,(L,D-\alpha ) =
0$, or to find factorization of the following type,
$$
L =(D - \alpha ) L_{1}, \qquad \mbox{ ord }L_{1} = n - 1
$$
\noindent the necessary and sufficient condition is
$\mbox{LRes}\,(L,D-\alpha )   = \mbox{RRes}\,(L^{*},D+\alpha )$, i.e.
$$
\left|
\begin{array}{ccccccc} a^*_n & a^*_{n-1} & \ldots & a^*_2 & a^*_1 &
 a^*_0 \\
1 & \alpha & \ldots & \left( {n-1 \atop n-3 } \right) \alpha^{(n-3)} &
\left( {n-1 \atop n-2 } \right) \alpha^{(n-2)} & \alpha^{(n-1)} \\
\ldots &\ldots & \ldots & \ldots & \ldots & \ldots & \ldots \\
0 & 0 & \ldots & 1 &  \alpha & \alpha' \\
0 & 0 & \ldots & 0 &  1 & \alpha
\end{array} \right| = 0.
$$
Thus we obtain the generalzed Ricati equation
$N^{*}(\alpha )$ of second kind and degree $(n-1)$:
$$
N^{*}(\alpha ) = L^{*}(\alpha ) + M^{*}(\alpha ),
$$
where
\begin{eqnarray}
L^{*}(\alpha ) \equiv \tau L = \sum_{r=0}^n \sum_{k=r}^n (-1)^{k} \left(
{k \atop r} \right)  a^{(k-r)}_{k} \alpha^k  = L^{*}_{1}(D+\alpha) .
\end{eqnarray}
Here $\tau$ is the conjugation operator, $L^{*}(\alpha )$ is
pseudo-characteristic Riccati equation of second kind and
\begin{eqnarray}
M^{*}(\alpha ) = \sum_{r=1}^n \sum_{k=r}^n (-1)^{k} \left( {k \atop r}
\right)  a^{(k-r)}_{k} M_{r-1}, \qquad
M_{r-1} = (D + \alpha )^{(r-1)} - \alpha ^{(r-1)},
\end{eqnarray}
$M^{*}(\alpha )$ is the reduced Riccati equation of second kind.

When $n=2$ the generalized Riccati equation of second kind have the form,

$$
N(\alpha ) = a_{2}\alpha ' + a_{2}\alpha ^{2} + (-a_{1}+2a_2')
\alpha  + a_{0}  - a_1' + a_{2}''= 0
$$
As a consequence of problem $1$ we have:

\medskip {\bf Problem 2.} Let us define the following equations,
\begin{eqnarray}
&\displaystyle
L_{1}y =\sum_{k=0}^n a_{k}D^{k}y = 0, \qquad a_{n}\neq  0, \quad  a_{k}\in
C^{k}_{I} ,& \\
&\displaystyle
L^{*}_{1}y \equiv \tau L_{1} =\sum_{r=0}^n \sum_{k=r}^n (-1)^{k}
\left( {k \atop r} \right)  a^{(k-r)}_{k}y^{k} =0.
\end{eqnarray}
Find the generalized equation of first kind,
$$
N_{1}(\alpha ) =\mbox{RRes}\, ( L_{1}, D - \alpha  )
$$
and the generalized Riccati equation of second kind,
$$
N^{*}_{2} (\alpha ) =\mbox{RRes}\,( L^{*}_{1}, D + \alpha  ) ).
$$

\medskip {\bf Algorithm 2.} We solve the problem $2$
by the procedure

$RICCATI(a,n,pp,x)$.

\begin{description}
\item[Input]{}
\item[$a$] is the array of coefficients of differential operator $L_{1}$;
\item[$n$] is the degree of differential operator $L_{1}$;
\item[$pp$] is the kind of seeking Riccati equation $(pp=1$ and $pp=2)$;
\item[$x$] is the independent variable.
\item[Output:]
The generalized Riccati equation of first kind $(pp=1)$ and
of second kind $(pp=2)$.
\item[R1:] If $pp=1$ then go to R2 else go to R3;
\item[R2:] $\ll L2 = (D - \alpha )y$;
\begin{description}
 \item[$b$] is the array of coefficients of operator $L2$;
 \item[$m:=1$];\quad
$N_{1}(\alpha ):=\mbox{DIFRESULT}(a,n,b,m,x)$; \quad  return
$N_{1}(\alpha ) \gg $;
\end{description}
\end{description}
\begin{description}
\item[R3:] $\ll L^{*}_{1} = \tau L_{1}$,\quad $\tau$ is operator of
conjugation,
$$
L^{*}_{1} = \sum_{r=0}^n \sum_{k=r}^n (-1)^{k} \left( {k \atop r} \right)
 a^{(k-r)}_{k}y^{k},
$$
\end{description}
\begin{description}
\item[$a^{*}$] is the array of coefficients of differential operator
$L^{*}_{1}$, \quad $L^{*}_{2}= (D + \alpha )y$;
\item[$b^{*}$] is the array of the coefficients of differential operator
$L^{*}_{2}$;
\item[$m:=1$];\quad
$N_{2}(\alpha ):=\mbox{DIFRESULT}(a^*,n, b^*,m, x)$;\quad  return
$N_{2}(\alpha ) \gg $;
\end{description}

\section{ ODE resolved by algebraic means}

{\bf 4.3.1.  Linear differential equations with exponential solutions
of type
$y = \exp (-\alpha x)$,\quad $ \alpha =
\mbox{const}$.}

The necessary and sufficient condition for existence of such solutions is
the pseudo-characteristic Riccati equation
of first kind $L(\alpha )$  and of second kind $L^{*}(\alpha )$
to have solution $\alpha = \mbox{const}$, i.e.
$\mbox{RRes}\,(L,D+\alpha)=0$
or $\mbox{RRes}\,(L^{*},D-\alpha ) = 0$.  Then the linear differential
equation
$Ly =\sum_{k=0}^n a_{k}D^{k}y$ have the form
$Ly = L_{2}(D + \alpha )y$,  where  $L_{2}$  is differential
operator,
$\mbox{ord}\,(L_{2})=n-1,$ and the differential equation have
the exponential solution of following type $y = \exp (-\alpha  x)$.
In the case of linear equation of second degree
$$
Ly = a_{2}(x)y'' + a_{1}(x)y' + a_{0}(x)y =0, \qquad a_{k}\neq 0 , \quad
a_{k}\in  C^{k}_{I}, \quad k=0\div 2 ,
$$
the problem is to find the coefficients of factrization
$\alpha _{1}$, $\alpha _{2}$ in
$Ly = (D - \alpha _{1})(D - \alpha _{2})y = 0$, where $\alpha_1 =
\mbox{const}$  and $\alpha_2 = \mbox{const}$.
i.e. we consider the classes of equations for which the associated
Riccati equation is of first type or of second type in terms of $\alpha $
have constant solutions.
\medskip {\bf Problem 3.} Given LEDE of degree $n$,
$$
Ly =\sum_{k=0}^n a_{k}D^{k}y =0 ,\qquad  a_{n}\neq 0, a_{k}\in  C^{k}_{I}.
$$
Find exponential solution  $y = \exp^(-\alpha x)$,
where $\alpha = \mbox{const.}$.

\medskip {\bf Algorithm 3.} In our program the problem 3 is solved by the
procedure
\begin{description}
\item[]  $LIVDIF(a,d,n,x)$.
\item[Input]
\item[$a$] is the array of coefficents of operator $L$;
\item[$d$] is nonhomogeneous part of equation $Ly$;
\item[$n$] is the degree of differential operator $L$;
\item[$x$] is the independent variable.
\item[Output]
If we find the particular solution of the following form
$y = \exp (-\alpha x)$, $\alpha = \mbox{const}$ and differential
equation is of degree two the coefficients of factorization
$\alpha_{1}$, $\alpha_{2}$ and
the fundamental system of solutions
$y_{1}(x)$, $y_{2}(x)$ are obtained. If the degree of LODE
is $n > 2$, then the degree of LODE is reduced by 1.
If the exponential solution does not exists the message is received.

\item[D1:] Algorithm 2 for finding the generalized Riccati equation
of first type $N_{1}(\alpha )$, \\
$N_{1}(\alpha ):= RICCATI(a,n,1,x)$;

\item[D2:] $L_{1}(\alpha ) =\sum_{k=0}^n a_{k}\alpha ^{k}$;
 $L_{1}(\alpha )$ - pseudo-characteristic equation of first kind;
\item[D3:] $M_{1}(\alpha )=N_{1}(\alpha )-L_{1}(\alpha )$;
 $M_{1}(\alpha )$ - reduced equation of first kind;
\item[D4:]  Subalgorithm for finding constant solution of algebraic
\\ equation $L_{1}(\alpha )$ in terms of $\alpha$. \\
{\bf \qquad Output:} constant solution $\alpha _{0}$, or \\ obtain
message that the solution of this type does not exists,
\item[D5:] If $\alpha _{0}$ exists and is solution of
$M_{1}(\alpha ) = 0$ then go to D11
 else go to D6;
\item[D6:] Algorithm 2 for finding the generalized Riccati equation of
second type $N_{2}(\alpha )$.\\
$N_{2}(\alpha ):= \mbox{RICCATI}(a,n,2,x)$;

\item[D7:]
\begin{eqnarray}
L_{2}(\alpha ) = \sum_{r=0}^{n} \sum_{k=r}^{n} (-1)^{k}
\left( {k\atop r}\right) a_{k}^{(k-r)}\alpha^k
\end{eqnarray}
($L_{2}(\alpha)$-pseudo-characteristic equation of second type)
\item[D8:]
\begin{equation}
M_{2}(\alpha)=N_{2}(\alpha) - L_{2}(\alpha)
\end{equation}
($M_{2}(\alpha)$-reduced Riccati equation of second type)
\item[D9:]
    Subalgorithm of finding constant solution of algebraic equation
$L_{2}(\alpha)$ in terms of $\alpha$
(Output- $\mbox{const.}$ the solution $\alpha_{0}$, or the message
that such a solution does not exists)
\item[D10:] If $\alpha _{0}$ exists and is solution of
$M_{2}(\alpha ) = 0$ then go to D11
 else go to D13;
\item[D11:] If $n=2$ then go to D12 else go to D13;
\item[D12:]
<< $\alpha_{2}=\alpha_{0};$
$\alpha{1}=-(a_{1}+\alpha_{2});$
\begin{equation}
 y_{1}=\exp(\int\alpha_{1}dx), \quad
  y_{2}=y_{1}\int\exp(\int(\alpha_{2}-\alpha_{1})dx)dx;
\end{equation}
Message  "The coefficients of factorization $\alpha_{1}$ and
$\alpha_{2}$ and fundamental solutions $y_{1}$ and $y_{2}$
are obtained " >>;
\item[D13:]
<< $ y_{1}=\exp(\int\alpha_{0}dx)$,
Message  "The degree of differential equation is reduced by 1";
New Lode := Sub($y=y_{1}\int zdx, Ly$);
 In equation
\begin{equation}
Ly =\sum_{k=0}^n a_{k}D^{k}y =0, \quad a_{n}\neq 0 ,
\quad a_{k}\in C^{k}_{I}
\end{equation}
we make the change of variables $y=y_{1}\int zdx$
where $z$ is new variable)
Return New Lode >>;
\item[D14:]
Message "The solution of exponential type does not
exists";
    Subalgorithm for finding of constant solution of algebraic equation
$L(\alpha)$ in terms of $\alpha$.
In our program this subalgorithm is realized by procedure
   $\mbox{HAREQ}(L(\alpha),\alpha,x)$;
Output:
$L(\alpha)$ is algebraic equation in term of $\alpha$.
$x$ is independen variable.
Output:
  We have $\mbox{const.}$ is the solution $\alpha_{0}$ of equation
$L(\alpha)$, or message, such a solution does not exists
  Example:
\begin{equation}
Ly=y'' + (8 + \sin(x)^{2})y' + 8\sin(x)^{2}y =0; \quad (')=d/dx
\end{equation}
In the array $a(n)$, $n=0\div 2$, we write the coefficients of the equation
\begin{equation}
a(2):=1; \quad a(1):=8 + \sin(x)^{2}; \quad a(0):=8\sin(x)^{2}; \quad
n:=2; \, d:=0;
\end{equation}
Using the procedure $\mbox{LIVDIF}(a,d,n,x)$; we seek particular
solution of the following type  $y = \exp^(-\alpha x)$,
where $\alpha = \mbox{const.}$, and if the differential equation is of
degree $n=2$ , then we find the coefficents of factorization
$\alpha_{1}$, $\alpha_{2}$ and the fundamental system of solutions
$y_{1}(x)$, $y_{2}(x)$, also the general solution.
Then we have the solution of differential equation of type
$y = \exp^(-8x)$, and the coefficents of factorization are
\begin{equation}
\alpha_{1}=8; \quad \alpha_{2}=\sin(x)^2,
\end{equation}
and due to the fact that the equation is of second degree we may found and
second solution
\begin{equation}
y_{2}=y_{1}\int\exp(\int(\alpha_{2}-\alpha_{1})dx)dx,
\end{equation}
and general solution
\begin{equation}
y=C1y_{1}+C2y_{2}.
\end{equation}

{\bf 4.3.2. LODE with polynomial solutions}
To exists in the set of solutions of the equation
\begin{equation}
Ly =\sum_{k=0}^n a_{k}D^{k}y =0, \quad a_{n}\neq 0 ,
\quad a_{k}\in C^{k}_{I}
\end{equation}
polynomial of degree $p\neq m$ the necessary and sufficent condition
is $\mbox{RRes}(L,D^{m+1})=0$, where
\begin{equation}
L =\sum_{k=0}^n a_{k}D^{k}, \quad D=d/dx ,
\quad D^{m+1}=d^{m+1}/dx,
\end{equation}
i.e. using the differential resultant of operators $L$ and
$D^{m+1}$ we find the degree of existing polynomial solution
of LODE and after that by metod of indefinite coefficents we find
the coefficents of this polynomial and if the degree
of LODE is $n=2$ we may obtain the general solution and when $n>2$ we
reduce by 1 the degree of LODE.

\medskip {\bf Problem 4.} Given LODE of degree $n$:
\begin{equation}
Ly =\sum_{k=0}^n a_{k}D^{k}y =0, \quad a_{n}\neq 0 ,
\quad a_{k}\in C^{k}_{I}
\end{equation}
find the polynomial solution.
\end{description}
\medskip

{\bf Algorithm 4.} To solve the problem 4 we apply the procedure

       POLDIF(a,n,x,hs);

\begin{description}
\item[Input]
\item[$a$] is the array of coefficents of LODE $Ly=0$,
\item[$n$] is the degree of differential operator $L$,
\item[$x$] is independent variable,
\item[$hs$] is the maximal degree of the polynomial solution,
\item[Output:]
\end{description}

If there exists the polynomial solution of degree $p\leq hs$, then
in the case of $n=2$ we find the general solution, and when
$n > 2$  the degree is reduced by 1. If there is no solution we have the
message that polynomial soluion does not exists.

\begin{description}
\item[P1:] $\quad i:=1$;
\item[P2:]   Algorithm 1 for finding the differential resultant
$\mbox{RRes}\,(L,D^{i})$;
\item[P3:]   If $\mbox{RRes}\,(L, D^{i}) = 0$, then go to P6 else go to
P4;
\item[P4:]   If $i>hs$ then go to P12 else go to P5;
\item[P5:] $\ll i:=i+1$;\quad go to \quad P2 $\gg $;
\item[P6:] P$:=i-1$;
\item[P7:]   Message ``$P$ is the degree of polynomial solution'';
\item[P8:]   Subalgorithm for finding the coefficents of the
polynomial solution ( Output- polynomial soluition)
\item[P9:]   If $n=2$ then go to P10 else go to P11;
\item[P10:]  Find the fundamental system of LODE.\\
 $\ll y_{1}=$ plynomial solution;
$$
y_{2}= y_{1}\int \exp \left(- \int a_{1}dx \right) y^{-2}_{1} dx
$$
Return \quad $y_{2},y_{1}\gg $;
\item[P11:] The degree of LODE is reduced by 1.\\
<< $ y_{1}=$ plynomial solution;\\
New Lode $:= \mbox{Sub}( y = y_{1}\int zdx, Ly)$, where
in equation
\begin{equation}
Ly =\sum_{k=0}^n a_{k}D^{k}y =0, \quad a_{n}\neq 0 ,
\quad a_{k}\in C^{k}_{I}
\end{equation}
we make the change of variables $y=y_{1}\int zdx$
where $z$ is the new variable\\
Return New Lode >>;
\item[P12:]   Message ``The finded polynomial solution of the
LODE '';
\end{description}

Subalgorithm for finding the coefficents of the polynomial solution
of the LODE.

In our program this subalgorithm is realized by the procedure
$DEPOLDIF(a,n,x,p)$;
\begin{description}
\item[Input:]
\item[$a$] is the array of coefficents of of the differential equation $y=0$
\item[$n$]  is the degree of differential operator $L$,
\item[$x$] is independent variable,
\item[$p$] is the degree of polynomila solution of $Ly=0$.

\item[Output:]
The coefficents of the polynomial solution of by the method of indefinite
coefficents.
\end{description}

\medskip
{\bf Example. }
$$
Ly = x^2 (\ln (x)-1)y'' - x y' + y =0, \qquad (') = d/dx
$$

 The array $a(n), n=0\div 2,$ contain the coefficents of the given LODE.

$$
a(2):= x^2(\ln (x)-1); \qquad a(1):=-x; \qquad a(0):=1; \quad n:=2;
\quad hs:=3;
$$
By using the procedure $POLDIF(a,n,x,hs)$ we find the fundamental system
of solution
$y1:=x; y2:=-\ln (x)$. Then the general solution is given by:
$y:=C1 x + C2\ln (x)$;

{\bf 4.3.3. The equations in terms of exact differentials.}
Let us define
 \begin{equation}
Ly =\sum_{k=0}^n a_{k}D^{k}y =0, \quad a_{n}\neq 0 ,
\quad a_{k}\in C^{k}_{I}, \quad D=d/dx
\end{equation}
This equation is expressed in terms of exact differentials of
there exists the following form
\begin{equation}
Ly =D\sum_{k=0}^{n-1} b_{k}D^{k}y =0, \quad D=d/dx,
\end{equation}
where
\begin{equation}
a_{n}=b_{n-1}(x), \quad a_{0}=b_{0}'(x),
\quad a_{k+1}=( b_{k}+b_{k+1}'(x)),
 \qquad (')=d/dx, \quad k=0\div (n-2).
\end{equation}
The necessary and sufficient condition the equation $Ly=0$ to be
in exact differential is,
\begin{equation}
\sum_{k=0}^n (-1)^{k}a_{k}^{k}(x) =0,
\end{equation}
If the equation
\begin{equation}
Ly =\sum_{k=0}^n a_{k}D^{k}y =0,
\end{equation}
is expressed in terms of exact differentials in the case  $n=2$ we may
find the general solution of this equation and when $n>2$ we may
find fist integral.
\medskip {\bf Problem 5.} Given LODE of degree $n$:
\begin{equation}
Ly =\sum_{k=0}^n a_{k}D^{k}y = f(x), \quad a_{n}\neq 0 ,
\quad a_{k}\in C^{k}_{I}
\end{equation}
Check is this solution of LODE in exact differentials and if the equation is
of this type find the general solution (when $n=2$) and
first integral (when $n>2$).

{\bf Algorithm 5.} To solve the problem 5 we apply
  the procedure

       EQDIF(a,n,dl,x);

\begin{description}
\item[Input:]
\item[$a$] is the array of coefficents of LODE
\item[$n$] is the degree of differential operator $L$,
\item[$dl$] is the nonhomogeneous part of given LODE
\item[$x$] is independent variable,
\item[Output:]
\end{description}
 Check is given equation in exact differentials and if it is true find
the general solution in the case $n=2$ and first integral
if $n>2$, and if it not the case,
message that the equation is not in exact differentials
\begin{description}
\item[E1:]
\begin{equation}
D:=\sum_{k=0}^n (-1)^{k}a_{k}^{k}(x) =0,
\end{equation}
\item[E2:] If $D=0$ then go to E4 else go to E3;
\item[E3:] Message:
   "This equation is not in exact differentials".
\item[E4:]
<< $b(n-1):=a(n)$;
For $k:=(n-2)$ Step (-1) Until 0 Do
\begin{equation}
b(k):=a(k+1)-b(k+1)',  \qquad (') = d/dx
\end{equation}
\item[E5:]
If $n=2$ Then Go To E6 Else Go To E7;
\item[E6:]
\begin{equation}
y=\exp(-\int a(1)dx) (c_{2} +
 \int(\int dl dx +C_{1})\exp(\int a(1)dx)dx);
\end{equation}
( $C_{1}$, $C_{2}$ - are constant of integration,
$y$ - general solution of LODE)
\item[E7:]
\begin{equation}
L_{2}y =D\sum_{k=0}^{n-1} b_{k}D^{k}y , \quad D=d/dx,
\end{equation}
\end{description}

{\bf Appendix 3: Differential resultant and algebra of commuting diffrential
operators.}

\begin{center}
{\bf Example 1}
\end{center}
\begin{eqnarray} \label{HHii}
H = \frac{1}{2} \left(p_1^2 + p_2^2 \right) +
                 \frac{1}{2} q_1q_2^2 + q_1^3 .
\end{eqnarray}
We associate to this Hamilonian (\ref{HHii}) the following
linear differential operators
\begin{eqnarray}
&& L=D^2 + q_1, \\
&& M=16 D^5 + 40 q_1 D^3 + 60 p_1 D^2 -
(120 q_1^2 + 25 q_2^2) D \nonumber \\ && -60q_1p_1 - 15 q_2 p_2,
\end{eqnarray}

This example show the abilities of the differntial resultant to
find the first inetgrals and algbraic curve associated to
given linear diferential operators. Using REDUCE 3.4 sintactis we
have
\begin{verbatim}
depend qq1,x; depend qq2,x; depend pp1,x; depend pp2,x;
a(0):=-60*qq1*pp1-15*qq2*pp2-zz;
a(1):=-120*qq1**2-25*qq2**2; a(2):=60*pp1;
a(3):=40*qq1; a(4):=0; a(5):=16;
b(0):=qq1-ll;  b(1):=0; b(2):=1;
b(3):=0;
out nik52kdv;
res52:=difresult(a,5,b,2,x)$
off mcd; res52:=num(res52)/den(res52);
write "denumerator:=",den(res52);
p1:=deg(res52,zz); write "p1:=",p1;
p2:=deg(res52,ll); write "p2:=",p2;
for i:=0:p1 do
<<
d1(i):=coeffn(res52,zz,i);
LET df(qq1,x,2)=-3*qq1**2-qq2**2/2;
LET df(qq2,x,2)=-qq1*qq2;
LET df(pp1,x)=-3*qq1**2-qq2**2/2;
LET df(pp2,x)=-qq1*qq2;
LET df(qq2,x,3)=df(-qq1*qq2,x);
LET df(qq2,x)=pp2,df(qq1,x)=pp1;
LET df(qq2,x,5)=df(-qq1*qq2,x,3);
LET df(pp1,x,2)=df(-3*qq1**2-qq2**2/2,x);
LET df(pp2,x,2)=df(-qq1*qq2,x);
LET df(pp2,x,4)=df(-qq1*qq2,x,3);

Determinant of resultant matrix

i-degree of zz:=0

j - degree of ll:=0

d2(0,0):=0

***********************************
i-degree of zz:=0

j - degree of ll:=1

              2    2            2      4
d2(0,1):=4*qq1 *qq2  - 8*qq1*pp2  + qq2  + 8*qq2*pp1*pp2

***********************************

i-degree of zz:=0

j - degree of ll:=2

                  3             2         2         2
d2(0,2):= - 32*qq1  - 16*qq1*qq2  - 16*pp1  - 16*pp2

***********************************

i-degree of zz:=0

j - degree of ll:=3

d2(0,3):=0

***********************************

i-degree of zz:=0

j - degree of ll:=4

d2(0,4):=0

***********************************

i-degree of zz:=0

j - degree of ll:=5

d2(0,5):=-256

***********************************

p3:=0

i-degree of zz:=1

j - degree of ll:=0

d2(1,0):=0

***********************************

p3:=0

i-degree of zz:=2

j - degree of ll:=0

d2(2,0):=1

***********************************
\end{verbatim}

The associated alegebraic curve has the form
(the results are obtaine dby differential resultant
\begin{eqnarray}
z^2 = -256\lambda^5 -32 E\lambda^2 + 8 K\lambda
\end{eqnarray}
where
\begin{eqnarray}
& &E = \frac{1}{2}\left(p_1^2 + p_2^2\right) +
q1^3 + \frac{1}{2} q_1 q_2^2, \\
& &K=q_2 p_1p_2-q_1p_2^2+q_2^4 + 1/2 q_1^2q_2^2.
\end{eqnarray}

\begin{center}
{\bf Example 2} (Fordy example \cite{Fordy}
\end{center}
\begin{verbatim}
a(0):=-zz; a(1):=-5*qq2**2; a(2):=15*pp1;
a(3):=15*qq1; a(4):=0; a(5):=9;
b(0):=-ll;  b(1):=qq1; b(2):=0;
b(3):=1;
out nik53sk;
res53:=difresult(a,5,b,3,x)$
off mcd; res53:=num(res53)/den(res53); write "denumerator:=",den(res53);
p1:=deg(res53,zz); write "p1:=",p1;
p2:=deg(res53,ll); write "p2:=",p2;
for i:=0:p1 do
<<
d1(i):=coeffn(res53,zz,i);
LET df(qq1,x,2)=-qq1**2/2-qq2**2/2;
LET df(qq2,x,2)=-qq1*qq2;
LET df(pp1,x)=-qq1**2/2-qq2**2/2;
LET df(pp2,x)=-qq1*qq2;
LET df(qq2,x,3)=df(-qq1*qq2,x);
LET df(qq2,x)=pp2,df(qq1,x)=pp1;
LET df(qq2,x,5)=df(-qq1*qq2,x,3);
LET df(pp1,x,2)=df(-qq1**2/2-qq2**2/2,x);
LET df(pp2,x,2)=df(-qq1*qq2,x);
LET df(pp2,x,4)=df(-qq1*qq2,x,3);

d1(i):=d1(i);
%write "d1(",i,"):=",d1(i);
%write "***********************************";
p3:=deg(d1(i),ll); write "p3:=",p3;
for j:=0:p3 do
<< d2(i,j):=coeffn(d1(i),ll,j);
write "i-degree of zz:=",i;
write "j - degree of ll:=",j;
write "d2(",i,",",j,"):=",d2(i,j);
write "***********************************" >> >>;


Determinant of resultant matrix

denumerator:=1


i-degree of zz:=0

j - degree of ll:=0

d2(0,0):=0

***********************************

i-degree of zz:=0

j - degree of ll:=1

                   4    2          2    4        2
d2(0,1):= - 9/4*qq1 *qq2  - 3/2*qq1 *qq2  - 9*qq1 *qq2*pp1*pp2

                   6        3                2    2
          - 1/4*qq2  - 3*qq2 *pp1*pp2 - 9*pp1 *pp2

***********************************

i-degree of zz:=0

j - degree of ll:=2

d2(0,2):=0

***********************************

i-degree of zz:=0

j - degree of ll:=3

               3             2         2         2
d2(0,3):=27*qq1  + 81*qq1*qq2  + 81*pp1  + 81*pp2

***********************************

i-degree of zz:=0

j - degree of ll:=4

d2(0,4):=0

***********************************

i-degree of zz:=0

j - degree of ll:=5

d2(0,5):=-729

***********************************

p3:=0

i-degree of zz:=1

j - degree of ll:=0

d2(1,0):=0

***********************************

p3:=0

i-degree of zz:=2

j - degree of ll:=0

d2(2,0):=0

***********************************

p3:=0

i-degree of zz:=3

j - degree of ll:=0

d2(3,0):=1

***********************************
\end{verbatim}
The associated algebraic curve has the form
(the results are obtained by differntial resultant)
\begin{eqnarray}
z^3 = 729\lambda^5 -162 E\lambda^3 + K^2\lambda,
\end{eqnarray}
where
\begin{eqnarray}
& &E = \frac{1}{2}\left(p_1^2 + p_2^2\right) +
\frac{1}{6}q1^3 + \frac{1}{2} q_1 q_2^2, \\
& &K=3p_1p_2 + \frac{3}{2}q_2 q_1^2 +\frac{1}{2} q_2 q_2^2,
\end{eqnarray}
\begin{verbatim}

print(`A program for calculation differential resultants`):
print(`written by N.A. Kostov, Z.T. Kostova 8 August 1994`):

#find the elements of series
procoff:=proc(s,k,m,b,x)
           local i,s1,s2,cr;
s1:=max(0,k-m); s2:=min(s,k);
if k < 0 or k > (m+s) then cr:=0 else
           cr:=0; for i from s1 to s2 do
     if (s-i)=0 then cr:=cr+binomial(s,i)*b(k-i) else
              cr:=cr+binomial(s,i)*diff(b(k-i),x$s-i);
                         fi;
                 od;
          fi;
RETURN(eval(cr))
end:

difresult:=proc(a,n,b,m,x)
        local i1,i2,i2,j2,i3,j3,k,s,l,p;
#****************************
#** Programm for finding   **
#** differential resultant **
#****************************
r:=array(1..n+m,1..n+m):
 for s from 0 to (m-1) do
    for k from 0 to (n+s) do
 ak(k,s):=procoff(s,k,n,a,x);
    od;
  od;
 for p from 0 to (n-1) do
    for l from 0 to (m+p) do
     bk(l,p):=procoff(p,l,m,b,x);
    od;
 od;
 for i1 from 1 to m do
     for j1 from 1 to (n+m) do
        if i1 > 1 and j1 < i1 then r[i1,j1]:=0 else
               r[i1,j1]:=ak(n+m-j1,m-i1);
        fi;
     od;
 od;
 for i2 from 1 to n do
        for j2 from 1 to (n+m) do
             if i2>1 and j2<i2 then r[m+i2,j2]:=0 else
                    r[m+i2,j2]:=bk(n+m-j2,n-i2);
             fi;
        od;
 od;
 print (`"Element of resultant matrix"`);
# for i3 from 1 to (n+m) do
#    for j3 from 1 to (n+m) do
#        print(`r:=`,r[i3,j3],i3,j3);
#    od;
# od;
with(linalg,det);
rres1:=det(r):
RETURN(rres1)
end:

#alias( dq1=dq1(x),q1=q1(x),q2=q2(x),dq2=dq2(x)):
#b(0):=dq1-z;
#b(1):=2*q1;  b(2):=0; b(3):=1;
#a(0):=-5*q2*dq2-60*q1*dq1-ll;
#a(1):=-120*q1**2-(35/4)*q2**2; a(2):=45*dq1;
#a(3):=30*q1; a(4):=0; a(5):=9;
#res3:=difresult(a,5,b,3,x);

#alias(UU=uu(x),WW=ww(x)):

#alias(UUx=diff(uu(x),x),UUxx=diff(uu(x),x,x),UUxxx=diff(uu(x),x,x,x)):

#alias(WWx=diff(ww(x),x),WWxx=diff(ww(x),x,x),WWxxx=diff(ww(x),x,x,x)):

#alias(WWxxxx=diff(ww(x),x,x,x,x),WWxxxxx=diff(ww(x),x,x,x,x,x)):

#b(0):=-Z+diff(ww(x),x); b(1):=2*WW; b(2):=0; b(3):=1;

#a(0):=20*WW*diff(ww(x),x)+10*diff(ww(x),x$3)-L;

#a(1):=20*WW**2+35*diff(ww(x),x$2); a(2):=45*diff(ww(x),x);

#a(3):=30*WW; a(4):=0; a(5):=9;

#difresult(a,5,b,3,x);


alias( DQ1=dq1(x),Q1=q1(x),Q2=q2(x),DQ2=dq2(x)):

alias(DQ1x=diff(dq1(x),x),DQ1xx=diff(dq1(x),x,x),DQ1xxx=diff(dq1(x),x,x,x)):

alias(DQ2x=diff(dq2(x),x),DQ2xx=diff(dq2(x),x,x),DQ2xxx=diff(dq2(x),x,x,x)):

alias(DQ1xxxx=diff(dq1(x),x,x,x,x),DQ1xxxxx=diff(dq1(x),x,x,x,x,x)):

alias(DQ2xxxx=diff(dq2(x),x,x,x,x),DQ2xxxxx=diff(dq2(x),x,x,x,x,x)):


alias(Q1x=diff(q1(x),x),Q1xx=diff(q1(x),x,x),Q1xxx=diff(q1(x),x,x,x)):

alias(Q2x=diff(q2(x),x),Q2xx=diff(q2(x),x,x),Q2xxx=diff(dq2(x),x,x,x)):

alias(Q1xxxx=diff(q1(x),x,x,x,x),Q1xxxxx=diff(q1(x),x,x,x,x,x)):

alias(Q2xxxx=diff(q2(x),x,x,x,x),Q2xxxxx=diff(dq2(x),x,x,x,x,x)):

#alias(Q2Nxxx=diff(Q2xx(x),x));

#alias(Q2xx=Q2xx(x));

DQ1:=Q1x:
DQ2:=Q2x:

b(0):=DQ1-Z;
b(1):=2*Q1;  b(2):=0; b(3):=1;
a(0):=-5*Q2*DQ2-60*Q1*DQ1-L;
a(1):=-120*Q1**2-(35/4)*Q2**2; a(2):=45*DQ1;
a(3):=30*Q1; a(4):=0; a(5):=9;
ppp:=difresult(a,5,b,3,x):

rrr:=expand(subs(DQ1=Q1x,Q1xx=-4*Q1^2-Q2^2/4,Q2Nxxx=Q2xxx,
Q2xx=-Q1*Q2/2,Q2xxx=-Q1x*Q2/2-Q1*Q2x/2,
Q1xxxxx=-24*Q1x*Q1xx-8*Q1*Q1xxx-Q2x*Q2xx*3/2-Q2*Q2xxx/2,
DQ2=Q2x,Q1xxx=-8*Q1*Q1x-Q2*Q2x/2,ppp)):

Pres:=coeffs(rrr,[L,Z],`tres`):


printdifres:=proc(sss)
       local rr;
for i from 1 to nops([tres]) do
  rr:=expand(Pres[i]):
   print(`number=`);
   print(i);
   print(`spectral paremeters=`);
   print(tres[i]);
   print(`element=`);
   print(rr);
od;
end:


print(`*******************************************************************`);


print(`spectral parameters`,-tres[2]*162);
EE:=expand(subs(Q2xx=-Q1*Q2/2,Q1xx=-4*Q1^2-Q2^2/4,Pres[2])):
EE:=expand(EE)/(81*2):
print(`energy is equal to res2`);
print(EE);

print(`*******************************************************************`);

print(`spectral parameters`,3*tres[4]/4);
K:=expand(subs(Q2xx=-Q1*Q2/2,Q1xx=-4*Q1^2-Q2^2/4,Pres[4])):
K:=-4*expand(K)/3:
print(`second integral is res4`);
print(K);

print(`*******************************************************************`);
print(`spectral parameters`,tres[7]);

res7:=expand(subs(Q2xx=-Q1*Q2/2,Q1xx=-4*Q1^2-Q2^2/4,
                  Q1xxx=-8*Q1*Q1x-Q2*Q2x/2,Q2xxx=-Q1x*Q2/2-Q1*Q2x/2,
            Pres[7])):
res7:=expand(res7):
print(`element res7=`);
print(res7);

print(`*******************************************************************`);

print(`spectral parameters`,tres[3]);
res3:=expand(subs(Q2xx=-Q1*Q2/2,Q1xx=-4*Q1^2-Q2^2/4,
                  Q1xxx=-8*Q1*Q1x-Q2*Q2x/2,Q2xxx=-Q1x*Q2/2-Q1*Q2x/2,
            Pres[3])):
res3:=expand(res3):
print(`element res3 =`);
print(res3);

print(`*******************************************************************`);

print(`spectral parameters`,tres[1]);
res1:=subs(Q2xx=-Q1*Q2/2,Q1xx=-4*Q1^2-Q2^2/4,
                  Q1xxx=-8*Q1*Q1x-Q2*Q2x/2,Q2xxx=-Q1x*Q2/2-Q1*Q2x/2,
            Pres[1]):
res1:=expand(res1):
print(`element res1=`);
print(res1);

print(`*******************************************************************`);

print(`spectral parameters`,tres[5]);
res5:=subs(Q2xx=-Q1*Q2/2,Q1xx=-4*Q1^2-Q2^2/4,
                  Q1xxx=-8*Q1*Q1x-Q2*Q2x/2,Q2xxx=-Q1x*Q2/2-Q1*Q2x/2,
            Pres[5]):
res5:=expand(res5):
print(`element res5=`);
print(res5);

print(`*******************************************************************`);

print(`*******************************************************************`);

print(`spectral parameters`,tres[8]);
res8:=subs(Q2xx=-Q1*Q2/2,Q1xx=-4*Q1^2-Q2^2/4,
                  Q1xxx=-8*Q1*Q1x-Q2*Q2x/2,Q2xxx=-Q1x*Q2/2-Q1*Q2x/2,
            Pres[8]):
res8:=expand(res8):
print(`element res8=`);
print(res8);

print(`*******************************************************************`);

print(`*******************************************************************`);

print(`spectral parameters`,tres[6]);
res6:=subs(Q2xx=-Q1*Q2/2,Q1xx=-4*Q1^2-Q2^2/4,
                  Q1xxx=-8*Q1*Q1x-Q2*Q2x/2,Q2xxx=-Q1x*Q2/2-Q1*Q2x/2,
            Pres[6]):
res6:=expand(res6):
print(`element res6=`);
print(res6);

print(`*******************************************************************`);




\end{verbatim}

\begin{verbatim}
alias(q[1]=q[1](x),q[2]=q[2](x)); alias(wp=wp(x));
alias(u=u(x));
Eq[1]:=diff(q[1],x$2)+q[2]^2/2+3*q[1]^2+a[0]*q[1]-a[1]/4:
Eq[2]:=q[2]^3*diff(q[2],x$2)+q[2]^4*q[1]+q[2]^4*a[0]/4+a[4]/4:
wp1:=(4*wp^3-g2*wp-g3)^(1/2):
wp2:=eval(diff(wp1,x)):
wp3:=eval(expand(subs(diff(wp,x)=wp1,wp2)));
F:=-6; B:=0; a[0]=0;  G:=-144;
for i from 1 to 2 do
SEq[i]:=eval(expand(subs(q[1]=F*wp+B,
q[2]=(G*wp^2+E)^(1/2),Eq[i]))):
SSEq[i]:=expand(subs(diff(wp,x)^2=(4*wp^3-g2*wp-g3),
       diff(wp,x$2)=wp3,expand(SEq[i]))):
#print(`SSEq(`,i,`):=`,SSEq[i]);
n[i]:=degree(SSEq[i],wp):
for j from 0 to n[i] do
Cowp[i,j]:=coeff(SSEq[i],wp,j):
#Cowp1[i,j]:=expand(subs(F=0,G=0,Cowp[i,j]));
od:
#FF1[i]:=seq(Cowp[i,i1]=0,i1=0..n[i]);
FF2[i]:=seq(Cowp[i,i3],i3=0..n[i]);
od:
for i1 from 1 to 2 do
for j1 from 0 to n[i1] do
#print(`Cowp(`,i1,`,`,j1,`):=`, Cowp[i1,j1]);
print(`Cowp(`,i1,`,`,j1,`):=`,subs(a[0]=0,E=36*g2, Cowp[i1,j1]));
od;
od;
#for i2 from 1 to 2 do
#for j2 from 0 to n[i2] do
#print(`Cowp1(`,i2,`,`,j2,`):=`,Cowp1[i2,j2]);
#od:
#od:
#Fs:={FF1[1],FF1[2]};
#sols:=solve(Fs,{F,A,B,G,D,E,C1^2,C2^2,gama,beta});
with(grobner);
Fg:=[FF2[1],FF2[2]];
#Fg1:=subs(eta^(-1)=eta1,C1^2=CC1,C2^2=CC2,Fg);
#gsolve(Fg);
#gbasis(Fg1,[A,b,F,G,D,E,CC1,CC2,r,s,sigma,gama,beta,eta,eta1,g2,g3],plex);
#gbasis(Fg1,[A,D,C1^2,C2^2,gama,beta],tdeg);
#gsolve(Fg1,[A,D,C1^2,C2^2,gama,beta]);
#A1:=expand(subs(A=(-6*s-sigma*D)/eta,B=A^2/F-F*g2/4,G=-eta*F/sigma,
#                E=D^2/G-G*g2/4,Cowp[1,3]));
#D1:=expand(subs(A=(-6*s-sigma*D)/eta,B=A^2/F-F*g2/4,G=-eta*F/sigma,
#                E=D^2/G-G*g2/4,Cowp[2,3]));
#C12:=solve(Cowp1[1,0],C1^2);
#Gama1:=solve(Cowp1[1,1],gama);
#Gama2:=solve(Cowp1[1,2],gama);
#AA:=solve(Cowp1[1,3],A);
#C22:=solve(Cowp1[2,0],C2^2);
#Be1:=solve(Cowp1[2,1],beta);
#Be2:=solve(Cowp1[2,2],beta);
#DD:=solve(Cowp1[2,3],D);

\end{verbatim}

{\bf Acknowledgments.} N.A.K. acknowledges support of contract
F--442/1994 of the Bulgaria Ministry of Science, Education and
Technology.

\end{document}